\documentstyle[12pt,epsfig]{article}                            

\newcommand{\sss}{\Sigma_e}
\newcommand{\SSS}{\Sigma}

\newcommand{\beq}{\begin{equation}} 
\newcommand{\eeq}{\end{equation}}   
\newcommand{\thee}{{\large $\frac{\theta}{2}$}}

\newcommand{\dds}{{\delta \SSS/\SSS}}
\newcommand{\dde}{{{\delta E}/{E}}}

\newcommand{\deel}{{$\frac{\delta E}{E}$}}
\newcommand{\dssl}{{$\frac{\delta \SSS}{\SSS}$}}

\newcommand{\htab}{\rule[-3mm]{0mm}{10mm}}
\newcommand{\hhtab}{\rule[-5mm]{0mm}{12mm}}

\newlength{\dinwidth}       
\newlength{\dinmargin}      
\begin{document}   

\begin{titlepage}  

\begin{flushleft}  
\noindent  
{\tt DESY 97-136     }\\ 
{\tt July 1997}       \\
\end{flushleft}    
\vspace*{3.0cm}    
\begin{center}     
\begin{Large} {\bf 
Some Properties of the Very High $Q^2$ \\
Events of HERA
}\end{Large} \end{center}   
\vspace{2.0cm}     
\begin{center}     
\begin{large}      
Ursula Bassler, Gregorio Bernardi  \\       
\end{large}
\end{center}       
\begin{center}     
     Laboratoire de Physique Nucl\'eaire et des Hautes Energies\\
     Universit\'e Paris 6-7, 4 Place Jussieu,    
75252 Paris, France\\
{\it e-mail: bassler@mail.desy.de; gregorio@mail.desy.de}     
\end{center}       
\vspace{2.5cm}     
\begin{abstract}   
\noindent  
The kinematic reconstruction of neutral current high $Q^2$ events at HERA
is discussed in detail
using as an example the recently published events of the H1 and ZEUS
collaborations at $Q^2 > $ 15000 GeV$^2$ and M $>$ 180 GeV, which are
more numerous than expected from Standard Model predictions.
Taking into account the complete information of these events,
the mass reconstruction is improved and the difference between the average
mass of the samples of the two experiments is reduced from 26$\pm$10 
GeV to 17$\pm$7 GeV,
but remains different enough to render unlikely an interpretation of the
excess observed by the two collaborations as
originating from the decay of a single narrow resonance.
\end{abstract}    
\end{titlepage}

\section{Introduction}      
At the electron-proton ($ep$) HERA collider, the study of 
DIS is performed
in an unique and optimal way up to values of 
the  squared momentum transfer $Q^2$ of $10^{5}$ GeV$^2$. 
Recently, the two HERA collaborations H1 and ZEUS
 have  reported~\cite{H1VHQ2,
ZEUSVHQ2} an excess of events 
 at {\it very high} $Q^2$ (defined in the following as $Q^2>$ 15000 GeV$^2$)
compared to standard model expectations. This observation has triggered
an important activity on ``Beyond the Standard Model'' theories
which might explain the effect~\cite{Doksh}. 
The most favoured solution involves
the production of a new resonance, which after its
decay is
kinematically similar to a standard deep inelastic scattering (DIS) event.

Indeed, in the naive quark-parton model (QPM), a parton carrying a fraction
$x$ of the momentum of the proton scatters elastically against
an electron and in the Standard Model, this interaction
is representing a t-channel scattering occuring via the exchange of a
gauge boson such as a photon or a $Z^{\circ}$. Beyond the Standard Model,
the naive QPM picture can represent the formation of an s-channel resonance 
(generically called ``leptoquark'') subsequently followed by a two-body
decay. While in the first interpretation the Bjorken $x$ variable is
one of the two  relevant variables to characterize the scattering,
 in the second case the invariant mass $M$ of the
system formed is the physical quantity of interest. This mass is
related to $x$  in the naive QPM by $M=\sqrt{xs}$, $s$ being the
squared center of mass energy. In ``real''  interactions, the quantum 
chromodynamics (QCD) effects do not modify  much this picture at 
large $Q^2$, since it
is characterized in DIS by the  dominance of
one electron + one partonic jet + one remnant jet in the final state,
and is thus similar to  the one in which a leptoquark
is produced and decays in 2 bodies.
So, when looking at 
high $Q^2$ deep inelastic
scattering (DIS) in an inclusive way it is possible to distinguish
between these two scenarios by comparing the event rates in different
$x$ or $M$ intervals: the smooth evolution of the DIS cross-section as
a function of $x$ can be opposed to the appearance of a sharp 
peak in the invariant mass distribution, characteristic of a leptoquark.

The HERA effect  is complicated by the fact that  the
invariant mass distributions of the  event samples 
of the two collaborations differ: at a mass greater
than 180 GeV, the 7 H1 events appear clustered around
a mass of 200 GeV,  while the 5  ZEUS events are broadly distributed 
between 191
and 253 GeV. It is the subject of this paper 
to understand how significant is this 
difference and to provide a combined mass spectrum using the complete
available information of these events in order 
to give  an interpretation of  
the effect.

At HERA  the kinematic reconstruction
does  not need  to rely  on  the scattered lepton only, since the 
most important  part of the 
hadronic system is visible in the almost $ 4 \pi$ detectors H1 and     
ZEUS. This redundancy  allows for   
     an experimental control of the systematic 
errors and  the radiative corrections,
hence to determine in a more precise way
the  usual DIS
kinematic variables $x,y$ and $Q^2$ which are defined as: \\[.4cm]
$ \hspace*{1cm} 
  x = {Q^2}/ ({2 P.q})  
   \hspace*{0.5cm}  \hspace*{1cm} 
y = ({P.q}) / ({P.k}) 
       \hspace*{0.5cm}  \hspace*{1cm} 
 Q^2= -(k-k')^2 = -q^2  = xys \hspace*{0.5cm}  $ \\[.4cm]
$P,k $  being  the 4-vectors of the incident proton and lepton,
 $k'$ the scattered lepton one.  

In this report we  briefly review in section 2 the methods used at HERA
to determine the kinematics of the high $Q^2$ events.
In section 3 we characterize the high $Q^2$ region and introduce a new method
which optimizes the kinematic reconstruction by determining the measurement
errors on an event by event basis. Section 4 is devoted to the detailed
study of the very high $Q^2$ HERA events  and  discusses 
the mass distributions obtained.

\section{Kinematic Reconstruction  at HERA}     

In order to introduce the kinematic methods used at HERA in
the high $Q^2$ region let us start by a few definitions.
The initial  electron 
and proton (beam)      
energies are labeled $E_{\circ}$ and $P_{\circ}$. 
 The energy and polar angle\footnote{The positive $z$       
axis is defined at HERA as the incident proton beam direction.}
 of the scattered electron (or positron) are 
$E$ and $\theta$. After identification of the scattered electron, we can
reconstruct  the following 
 independent hadronic quantities: $\Sigma$, obtained      
as the sum of the scalar quantities $E_h-p_{z,h}$ associated to
each particle      
belonging  to the hadronic final state,  $p_{T,h}$
as its total transverse momentum and define the hadronic inclusive angle 
$\gamma$:
\begin{equation}   
 \Sigma = \sum_{h} (E_h-p_{z,h})    
 \hspace*{1cm}
     p_{T,h}   = \sqrt{(\sum_{h} p_{x,h})^2+(\sum_{h} p_{y,h})^2 }    
 \hspace*{1cm}
    \tan\frac{\gamma}{2} = \frac{\Sigma}{ p_{T,h}}      
\end{equation}     
$E_h,p_{x,h},p_{y,h},p_{z,h}$ are the four-momentum vector components   
of every energy deposit in the calorimeter originating
 from any hadronic final state particle. 
The corresponding quantities for the scattered electron are
\begin{equation}   
 \sss=E~(1-\cos{\theta})
\hspace*{2cm}   p_{T,e}=E\sin{\theta} \hspace*{1.2cm} \mbox{i.e.}
\hspace*{0.7cm}  \tan\frac{\theta}{2} = \frac{\Sigma_e}{ p_{T,e}}
\end{equation}     
   
Out of these variables, it is possible to write the 
four methods which have been used to determine the invariant mass of
the very high $Q^2$ events (the formulae
are given in the appendix).

\noindent
- The electron only method ($e$) based on E and $\theta$. \\
- The hadrons only method ($h$) based on $\Sigma$ and $p_{T,h}$~\cite{jb}.\\
- The double angle method (DA) based on  $\theta$ and 
        $\gamma$~\cite{stan}.\\
- The Sigma method ($\SSS$) based on E, $\theta$ and $\SSS$~\cite{bb}.

The $e$ method is  precise  at high $y$ but becomes less precise in $x,y$
or $M$ at low
$y$. The $h$ method
is the only one available for  charged current events, 
but is  less precise
than the three others in neutral currents, so it will be ignored
in the following. The DA method is  precise at high
$Q^2$ both
at high and low $y$.
 The $\SSS$ method is similar to the $e$ method at high $y$ and
to the DA at low $y$ apart from $Q^2$ which is less precise. However 
it is insensitive to initial state QED radiation (ISR)
in M, $y$ and $Q^2$. See \cite{bb2} for a detailed
discussion of the properties of these methods and \cite{wolf} for their
behaviour in presence of ISR.

It should be stressed that  the quality of any kinematic reconstruction
depends mainly on the precision achieved on the observables,
such as E, $\theta$, $\Sigma$ or $\gamma$. It is one of the  major tasks
of the experiments to obtain precise and unbiased quantities,
and several techniques have been developped in the HERA structure function
analyses to achieve these goals. We refer to the original publications for a 
description of these studies~\cite{HZ93-94}, which discuss for instance
how the electron energy is calibrated, or how the hadronic final state is
measured at low $y$, when the hadronic jets are in the vicinity of the beam
pipe.

\section{Treatment at High $Q^2$}

At {\it high} $Q^2$ (defined in the following as $Q^2>$ 2500 GeV$^2$)
the kinematic reconstruction is  in general  more precise
than at low $Q^2$ and indeed the differences between the results 
obtained with the methods seen above are small.
However since the number of events drops rapidly with increasing $Q^2$
it is still crucial  to optimize the reconstruction by making 
a full use of all the observables of these events.

The improvements on the reconstruction at high $Q^2$ come mainly
from the better measurement of the hadronic final state.
This is due to the fact that i) the individual hadron energy is on average
greater; ii) the losses in the beam pipe and in the material in front of
the calorimeter  are in proportion smaller than at
low $Q^2$, i.e. $\SSS$, $p_{T,h}$ and $\gamma$ 
are less affected by these losses;
iii) the hadronic final state displays more often a (single) collimated
 jet configuration.
This is due both to kinematics (events are in average at higher $x$),
and to the smaller gluon radiation and power corrections ($\alpha_S$ is 
smaller). 

All these characteristics enable a precise measurement of the hadronic
angle, and thus are particularly favourable to the DA method
which indeed becomes more accurate with increasing $Q^2$. This method
has however the drawback to be 
very sensitive to ISR
and  it is thus difficult
to apply to  a small number of events. 
For instance, the ISR of a 2.75~GeV photon, if not taken 
into account produces a shift
of 10\% on the reconstructed mass and  20\% on $Q^2$.
In order to overcome this drawback we introduce a new method 
which, by making use of the 
precision on $\gamma$, estimates {\it from the data} on an event by event
basis the shift in the measurement of 
 $E$ and $\SSS$, denoted in the following $\delta X/X \equiv 
(X_{true}-X_{reconstructed})/X_{reconstructed})$  with $X=E$ or $\SSS$.
These shifts allow
i)~the $\SSS$ method to be corrected; ii) the presence of an undetected
initial state photon (down to an energy of about 2 GeV) to be identified
with high efficiency, and therefore to correct for it.

\subsection{The  $\omega$ Method}
Let us  assume that there is no ISR (the specific treatment of ISR
is discussed in the next section, although the two steps are not
separated in the procedure), and that $\theta$ and $\gamma$
are precisely measured (implying 
 ${\delta \SSS}/{\SSS}=
{\delta p_{T,h}}/{p_{T,h}}$).  From energy momentum
conservation the two following equations can be derived:
\begin{eqnarray}
(1-y_e) \ \frac {\delta E}{E} + y_h \ \frac{\delta \SSS}{\SSS} = 
y_e - y_h
\\
-p_{T,e} \ \frac{\delta E}{E} + p_{T,h} \ \frac{\delta \SSS}{\SSS} = 
p_{T,e} - p_{T,h}
\end{eqnarray}
Under these assumptions  ${\delta E}/{E}$ and  ${\delta \SSS}/{\SSS}$
are determined on an event by event basis.
 Varying $\theta$ and $\gamma$ within their
typical errors\footnote{At very high $Q^2$, the $\gamma$
resolution improves to 30 mrad. We  use here
the r.m.s., rather than the standard deviation obtained from
a gaussian fit to the distribution,
to take into account non-gaussian tails which contribute
in the propagation of the systematic errors.}
 (5 and 40 mrad (r.m.s.)
respectively, at high $Q^2$) we can obtain
the errors arising from the angular measurements on
${\delta E}/{E}$ and  ${\delta \SSS}/{\SSS}$.
A more direct way to see the uncertainties arising from this
determination is illustrated  in fig.1, in which the error
on $E$ and $\SSS$ reconstructed  using equations 3 and 4
are compared to their ``true'' value, which is known for the
reconstructed simulated events.
For a given event, the error on $({\delta E}/{E})_{rec}$, which is defined as
$\Delta({\delta E}/{E})\equiv ({\delta E}/{E})_{rec} -
({\delta E}/{E})_{true}$, 
is of the same order as
the r.m.s. (denoted $<\dde>$) of the  $\dde$ distribution obtained from
a large sample (fig.1b),  showing that its use 
will not bring an improvement on an event by event basis. On the other side,
the
relative error on $(\dds)_{rec}$ is smaller, about 30\% of $<\dds>$ ,
and diminishes at high $y$ (compare
fig.1e to 1f, and 1b to 1c)
implying that correcting $\SSS$
 will provide a better kinematic measurement, in particular at high $y$.
\begin{figure}[h]                                                           
\begin{center}                                                                 
\epsfig{file=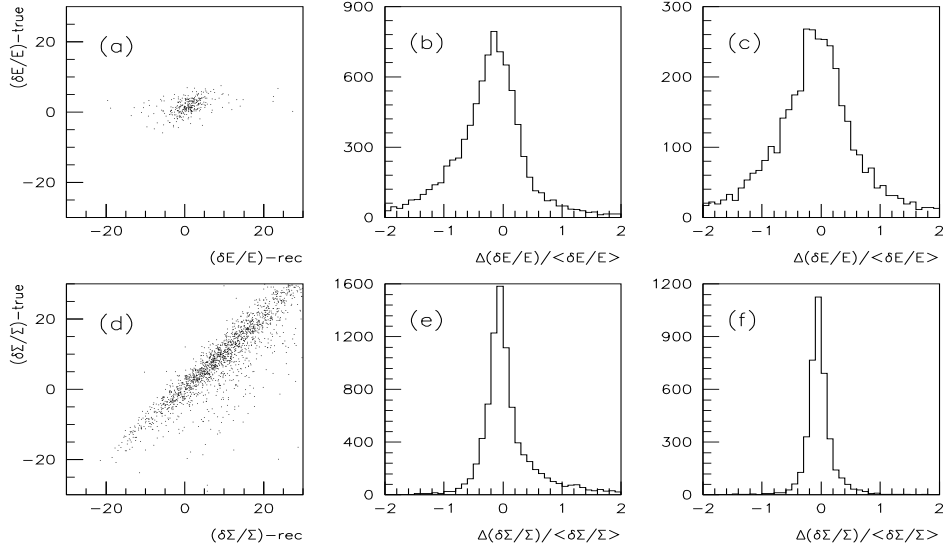,width=13cm,
height=7.2cm,bbllx=30pt,bblly=240pt,bburx=530pt,bbury=580pt} 
\end{center}    
\vspace*{-0.7cm}                                                               
\caption[]{\label{fig.4}
\sl { a) shows the correlation between  true 
reconstruction errors on $E$ and those obtained from eqs. 3 and 4,
on  a full simulation$^3$ 
in the H1 detector of high $Q^2$ events;
b) shows the projection of this correlation; c) shows the same projection
restricted to $ y>0.4 $. 
We use
$\Delta(\frac{\delta E}{E})\equiv (\frac{\delta E}{E})_{rec} -
(\frac{\delta E}{E})_{true}$, 
and $<\frac{\delta E}{E}>\equiv$ r.m.s. of
the $\frac{\delta E}{E}$ distribution = 3\%.
d,e,f)  show the same
plots but for $\SSS$ instead of $E$, 
with $<\frac{\delta \SSS}{\SSS}>$ = 10\%.}
}
\end{figure} 

 The $\omega$ kinematic variables
are thus derived from the $\Sigma$ ones by including the effect of
$\dds$, i.e.  
\begin{equation}
y_{\omega}\equiv\frac{\SSS + \delta \SSS} {\SSS + \delta \SSS + \SSS_e} 
\hspace*{2cm} Q^2_{\omega}\equiv\frac {p_{T,e}^2}{1 -y_{\omega}} 
\end{equation}
The comparison of the $\omega$ and $\SSS$ 
reconstruction of M, $y$ and $Q^2$ for high $Q^2$ events fully 
simulated\footnote
{In figures 1 and 2, 
the radiative processes of the DJANGO~\cite{django} program
which is used to generate the events 
have been turned off. The complete simulation, based on DJANGO,
ARIADNE~\cite{cdm} and GEANT~\cite{geant}
as used in the H1 simulation program, is further described
in~\cite{H1VHQ2}.}
in the H1 detector is shown in fig.2. 
The improvement obtained by the recalibration
of $\SSS$ is clearly visible, allowing this method 
to be compared favourably to the original ones.
\begin{figure}[htb]                                                           
\begin{center}                                                                 
\epsfig{file=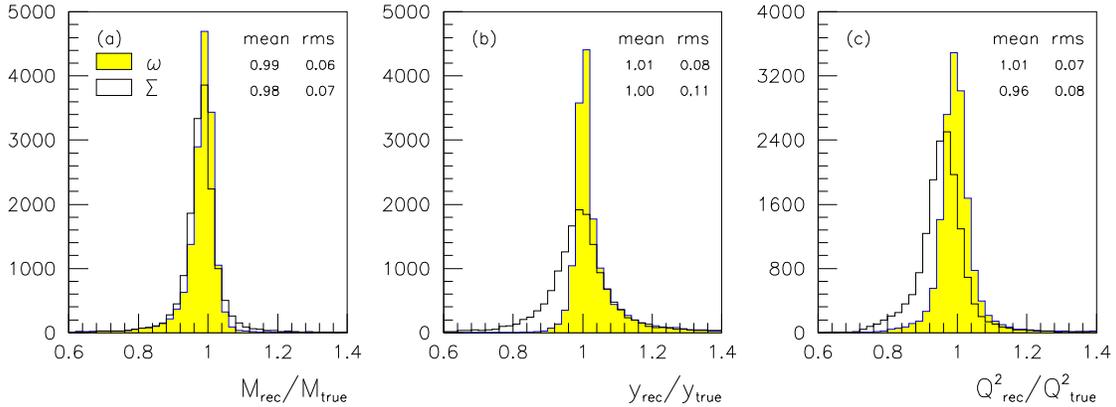,width=15cm,
  bbllx=35pt,bblly=320pt,bburx=525pt,bbury=505pt} 
\end{center}                                                                   
\caption[]{\label{fig.5}
\sl { The improvements on the kinematic reconstruction 
at high $Q^2$, when using the $\omega$ method compared to the $\SSS$ method,
are shown for the mass $M$ (2a), $y (2b)$ and $Q^2 (2c)$. 
}}
\end{figure}        
This comparison is shown in fig.3 for high $Q^2$ events 
(in all the high $Q^2$ section, an $E-p_z$ cut against hard initial
state radiation is applied, see next section)
both at high $y$ and at low $y$.
At high $y$ $(y_e>0.4$),  the $e$ and the $\omega$ methods 
are comparable  in $M$ and $y$
and slightly better than the DA one (fig.3a,b).
In $Q^2$,
the DA displays the narrowest peak both at high and low   $y$ (fig.3c,f), 
but  its high sensitivity to ISR
induces  tails in the distribution which renders its r.m.s. larger than the
$\omega$ and $e$ ones.
At low $y$ ($y_e<0.25)$ the $e$ method has a relatively
poor resolution in M, much worse
than the $\omega$  one which is slightly better than the DA one. 
In conclusion, for  the high $Q^2$ events the $\omega$ method 
is similar or slightly better than
 the $e$ and DA method.
\begin{figure}[htb]                                                           
\begin{center}                                                                 
\epsfig{file=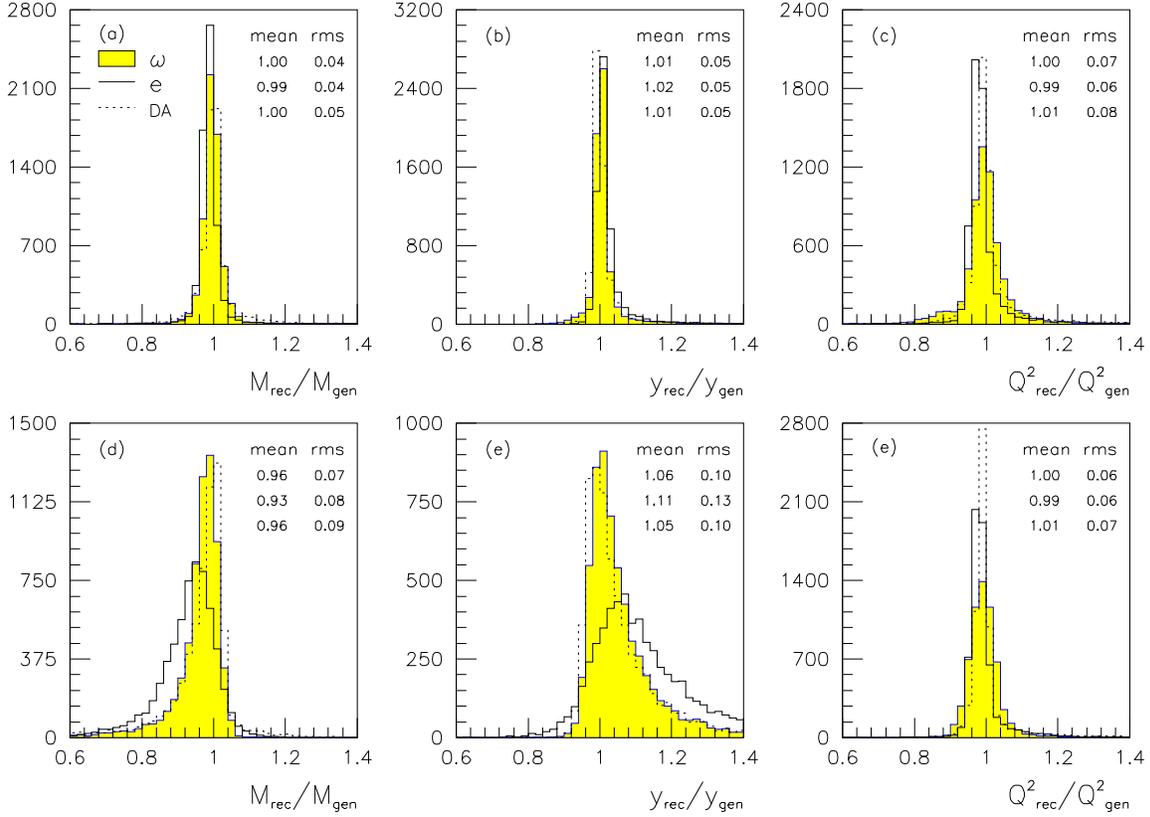,width=15cm,
  bbllx=50pt,bblly=250pt,bburx=525pt,bbury=580pt} 
\end{center}                                                                   
\caption[]{\label{fig.6}
\sl { Comparison on high $Q^2$ events,  at high $y$ 
($y_e>0.4$) of the $\omega$, DA and $e$ reconstruction of the mass (3a),
$y$ (3b) and $Q^2$ (3c). The same comparisons are made at low $y$ ($y_e<0.25$) 
in fig. 3d,e,f.}
}
\end{figure}

\subsection{Treatment  of QED Initial State Radiation}
The influence of ISR is taken into account in the simulation programs used
by the HERA experiments, but on a small number of events it
is difficult to control the migration due to an unseen ISR photon
when using the $e$ and DA methods. Both H1 and ZEUS use an 
experimental cut on the total $E-p_z$ of the event ($\SSS_{he}$)
or equivalently on the normalized $E-p_z$ called hereafter
$\sigma_{he}$\footnote{The normalized E-p$_z$ for the electron and for the
hadronic final state satisfy: $\sigma_e=1-y_e$;  $\sigma_h=y_h$.}
and defined as
\begin{equation}
 \sigma_{he} \equiv \frac{\SSS_{he}}{ 2E_{\circ}} 
  \equiv \frac{\SSS+\SSS_e} { 2E_{\circ}} 
\equiv \sigma_h + \sigma_e  
\end{equation}
The $E-p_z$ cuts used in H1 and ZEUS  correspond to  good approximation
to  $\sigma_{he}~>~0.75$ and  prevent an  ISR photon
from  carrying away more than about 25\% of the incident electron energy.
Such a photon can induce a large shift to the reconstructed mass~\cite{wolf} of
$1-\sqrt{(0.75  y_e ) / (y_e-0.25)}$ for $M_e$ (i.e. $-18\%$ at $y_e=0.75$) 
and of 25\% on $M_{DA}$ at any $y$. On $M_{\SSS}$ the shift is zero since 
the method is  independent of colinear ISR for  M$,y$ and $Q^2$.
A more stringent cut on $\sigma_{he}$ is difficult to implement due to the
experimental resolutions on $y_h$ and $y_e$. 

The $\omega$ method 
allows  ISR  to be identified
 with high efficiency for $\sigma_{he}$ as high
as $\sim0.93$. It makes use of the simple fact that depending on the origin
of the $\sigma_{he}$ shift (which comes mainly either from ISR or from
hadronic miscalibration), the errors obtained with the $\omega$ method
have a completely different pattern: for example if
the observed $\sigma_{he}$ is 1$-z$, equations 3 and 4 will give
for  the ISR case
$\delta E/E=z$ and $\delta \SSS/\SSS=z$,  assuming negligible
detector smearing,
while for the hadronic miscalibration case, assuming negligible
error on the electron energy, we will get 
$\delta E /E=0$ 
and $\delta \SSS/\SSS={z}/{y_h}$. 
These two  different types of pattern allow ISR  to be identified even in 
the presence of detector smearing, as studied on a complete  
simulation in the H1 detector, 
 with typically 85\% efficiency at high
$Q^2$ for photon energies greater than 2 GeV. The exact conditions for 
ISR identification  of a
2 GeV photons is $\delta E/E$ and $\delta \SSS/\SSS>7.3\%$.
However, to take into account the detector smearing
we use instead:
\begin{equation}   
\frac{\delta E}{E}>5\%      \hspace*{0.5cm} \mbox{and} \hspace*{0.5cm}  
\frac{\delta \SSS}{\SSS}>5\% \hspace*{0.5cm} \mbox{and} \hspace*{0.5cm}  
 \frac{\delta E}{E}+\frac{\delta \SSS}{\SSS}>15\%
\end{equation}   
These  can be slightly varied depending on 
the ISR identification efficiency requested. 
These conditions will ``misidentify'' {high $Q^2$} non-radiative 
events in less than 3\% of the cases in the H1 detector, implying that
in the total  sample of ISR identified
events, the fraction of genuine ISR events  is higher than the
fraction of non-radiative events.
This fraction increases with increasing $Q^2$ since
the hadronic angle becomes more precise, i.e. the ISR identification
becomes more efficient. 
In case of ISR identification, the equations 3 and 4 are solved again
after recalculating $y_e$ and $y_h$ using 
$\sigma_{he} \cdot 2E_{\circ}$ instead of $2E_{\circ}$.
 The improvement
obtained can be visualized in fig.4 which shows the same distributions
as in fig.3 but on the sample of DIS high $Q^2$ 
events having radiated an ISR photon
of 2 GeV or more. The dramatic improvement underlines the importance of
controlling experimentally the   radiation effect. 
Furthermore, the soft ISR photons which cannot be identified
below 2 GeV have a small
effect on $M_{\omega}$ ($<2\%$) but still a sizeable one  on $M_{DA}$,
up to about  7\%,  or 4\% on $M_e$.

\begin{figure}[htb]                                                           
\begin{center}                                                                 
\epsfig{file=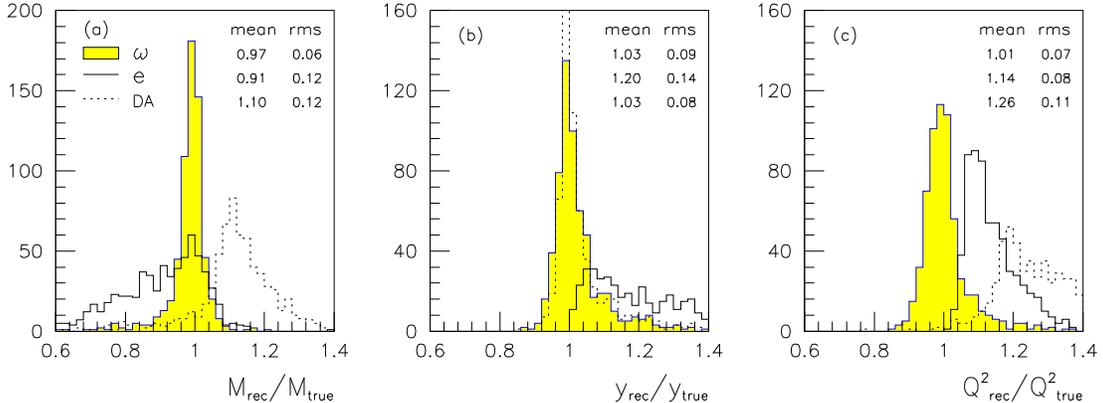,width=15cm,
  bbllx=35pt,bblly=310pt,bburx=525pt,bbury=500pt} 
\end{center}                                                                   
\caption[]{\label{fig.7}
\sl {  
Comparison  of the $\omega$,$e$ and DA  reconstruction of the mass (4a), 
$y$ (4b) and $Q^2$ (4c) for high $Q^2$ radiative events in which  
$E^{ISR}_{\gamma}> 2$ GeV (and $\sigma_{he}>0.75$). Note the similar
behaviour of $y_{\omega}$ and $y_{DA}$, since $y_{DA}$ is ISR independent.
In $M$ and $Q^2$ the improvements brought by the $\omega$ reconstruction
method
are remarkable.}
 }
\end{figure}        

\section{The Very High $Q^2$ Events at HERA}

Let us remind some characteristics of the  excess  of events at high $Q^2$
observed  by the H1 and ZEUS collaborations in the
data collected from  1994 to 1996~\cite{H1VHQ2,ZEUSVHQ2}. 
At $Q^2>15000$ GeV$^2$, 24 events have been observed
for an expectation of 13.4$\pm$1.0, representing a probability
of 0.0074~\cite{h1zeus}. 
Furthermore some of these events exhibit the additional
characteristics either to cluster around a rather precise value of
the invariant mass of the produced system
(M$_e=200\pm 12.5$ GeV, for 7 events out of the 12 of H1, while 
$1.0 \pm 0.2$ are expected), 
or to lie at very high invariant mass (M$_{DA}>223$ GeV and
$y_{DA}>0.25$, 
for 4 events out of the 12 of ZEUS, while $0.9\pm 0.2$ are expected).
Actually the messages of these  two observations could be different
as already pointed out~\cite{drees,ellis} and we will try here to explore
further this point with the use of our new kinematic tool. 

Among the obvious possible reasons for the difference are the influence
of   ISR which can strongly distort a distribution
based on a small number of events, or the effect of a specific miscalibration.
The $\omega$ method having been devised to respond to these problems, 
we can now study in more details the 11  extreme events mentioned 
above for which the complete kinematic information  has been 
published by the two collaborations.
We add to this sample  2 events\footnote{For one of these 2 events
(called Z--6 in table 1) only the 
DA variables
are available (from fig.1 of ref.~\cite{ZEUSVHQ2}).
The additional  kinematic properties 
were deduced from the DA values using the average $E-p_z$
and systematic shifts between $e$ and DA
variables of the 5 other ZEUS published events.
This assumption has a negligible influence on
the conclusions drawn below.
}
from ZEUS in order to include all
events which satisfy the
condition $Q^2>15000$ GeV$^2$ and M$>180$ GeV, since we want to test
the significance of the H1 event 
clustering at masses around 200 GeV over the complete very high $Q^2$ HERA
sample.

\subsection{Kinematic Properties}

In table 1 are given the M, $y$  and $Q^2$ 
of the 13 events at very high $Q^2$
and high mass reconstructed with the four different methods discussed above.
The values of M, $y$ and $Q^2$ are identical to those published
for the $e$ and  DA methods.  The  $\SSS$ values have  been released recently
by H1~\cite{GREGQ2} and can be computed straightforwardly for the ZEUS events.
The errors are also reproduced exactly except
 for those of the  $e$ method to which we conservatively added quadratically
an error of 1.5\% (i.e. half of the absolute energy scale uncertainty of 3\%
given by the two experiments)
to account for potential miscalibrations
between the different regions of the detectors.
As we will see below
the overall energy calibration can be checked to be well  under control.

For the errors not published by  the collaborations ($\omega$, $\SSS$ for 
H1 and ZEUS,
and DA for H1) we used the following prescription, which was checked to be
consistent with the other published errors:
They  are obtained here using a possible  error of 
$\pm 5$ mrad and $\pm 30$ mrad for the electron  and hadronic angle
respectively. For the error on the electron (hadronic) 
energy we used $\pm 3\%$ ($\pm 4\%$)
for H1 and $\pm 5\%$ ($\pm 4\%$) for ZEUS, which includes for the 
electron case both 
resolution effects, dead material corrections and the 1.5\%
overall error just mentioned. These two values for the electron
energy error have been checked on
the data (see below). Additional sizeable errors 
as published  by the collaborations, due for instance to special energy
corrections,  have been taken into account in two events (H--3,Z--1).

For the H1 events the errors of an event are similar for each
method, with
a slight advantage for the $e$ method at $y$ above 0.5. For the ZEUS events
the DA errors are the smallest, since the error on the electron energy
which is somewhat larger than in the H1 case has an influence on
the 3 other methods.
This also explains why the ZEUS collaboration chooses the DA method
rather than the $e$ method favoured by H1. 

Also mentioned is the result of the $\omega$ ISR identification, 
and we can see that
2 events (H--5 and Z--4) are classified as radiative, and have been
corrected accordingly.
These 2  events indeed show the  characteristics of a radiative event, i.e.
there exists 
a value of $z \equiv  E^{ISR}_{\gamma}/E_{\circ}$ for which the following
equations, which hold exactly if there were no detector effects, are verified
in a good approximation:
\begin{equation}   
   {{\delta E}/{E}} \simeq z \hspace*{1cm} 
   {{\delta \SSS}/{\SSS}} \simeq z 
\end{equation}   
\begin{equation}   
Q^2_{\SSS}/Q^2_{e} \simeq 1-z \hspace*{1cm} Q^2_{e}/Q^2_{DA} \simeq 1-z
\end{equation}   
\begin{equation}   
 M_{\SSS} / M_{DA} \simeq 1-z \hspace*{1cm}
 x_{\SSS} / x_{DA} \simeq (1-z)^2
\end{equation}   
\begin{equation}   
 y_{DA}\simeq y_{\SSS} \hspace*{1cm} \sigma_{he}  \simeq 1-z
\end{equation}   
Indeed in the case
of H--5 a photon with an energy (2.9 GeV)consistent with the  $z$ derived
from the previous equations 
is observed in the special photon calorimeter located
close to the beam pipe and designed to
measure photons emitted collinearly to the incident electron, such as
ISR photons.
 The ISR effect on the mass is sizeable, since $M_{DA}/M_{\omega}$ =
1.08 for H--5 and 1.11 for Z--4. Both values are 
consistent with the determined
$\delta E / E $ and $\delta \SSS / \SSS$ of these events (see tab.1).

The reconstructed kinematics are consistent between the different
methods except for the
2 radiative events and for the events displaying a significant difference
between $M_e$ and $M_{DA}$ (events H--7,Z--1,Z--5), which indeed
have the largest $\delta E / E $ and/or $\delta \SSS / \SSS$.

On these small samples of events, and after ISR corrections, the averages
of the~absolute value of the 
error on the hadronic energy are similar for H1 and ZEUS 
($<|\frac{\delta\SSS}
{\SSS}|>$= 6\% compared to  5\%),  while on the
electron energy the  H1 average error is smaller: $<|\frac{\delta E}
{E}|>$=3\% compared to 5\% for ZEUS.

The uncertainty on the 
electromagnetic absolute energy scale can be estimated {\it from the data}
using $<\frac{\delta E}{E}>$ (the 2 radiative events are excluded
from these means) which gives $+1.1\%\pm 1.5\%$ for H1 and
$+2.1\%\pm 3\%$ for ZEUS, both values in good agreement with the
value of 3\% given by the collaborations. 
The uncertainty on the 
hadronic absolute energy is obtained similarly
and gives $-5.4\%\pm 3\%$ for H1 and
$+3.3\%\pm 4\%$ for ZEUS, also  in  acceptable agreement with the
values quoted by the  collaborations. 

After all these consistency checks, the
use of the $\omega$ method
in the two  samples will now allow a consistent 
mass distribution to be derived from the very high $Q^2$ events.

\begin{table}
\begin{small}
\centering
\begin{tabular}{|c||c||c|c|c|c||c|c|c|c||c|c|c|c|}
\hline 
{\bf Evt}     & {\tiny  $\delta  E / E $ }  & 
$ M_{\omega} $& $ M_{DA}$&$  M_{e}    $&$ M_{\SSS}$&
$ y_{\omega} $& $ y_{DA}$&$  y_{e}    $&$ y_{\SSS}$&
$ Q^2_{\omega}$&$ Q^2_{DA}$&$  Q^2_e   $&$Q^2_{\SSS} $\\
$\sigma_{he} $&  {\tiny $\delta \SSS / \SSS $ }      
&{\tiny $ \delta M_{\omega}  $}&{\tiny $ \delta M_{DA}  $}
&{\tiny $ \delta M_e         $}&{\tiny $ \delta M_{\SSS}$}
&{\tiny $ \delta y_{\omega}  $}&{\tiny $ \delta y_{DA}  $}
&{\tiny $ \delta y_e         $}&{\tiny $ \delta y_{\SSS}$}
&{\tiny $ \delta Q^2_{\omega}$}&{\tiny $ \delta Q^2_{DA}$}
&{\tiny $ \delta Q^2_{e}     $}&{\tiny $ \delta Q^2_{\SSS} $}\\
\hline
\hline
 {\bf H--1}  &$ +.01    $&
       197 &       198 &       196 &       196 &
      .435 &      .434 &      .439 &      .443 &
      16.8 &      17.1 &      17.0 &      17.1 \\
1.01      & $ -.03    $&
   7 &   7 &   6 &   7 &
.016 &.016 &.016 &.032 &
 0.9  & 0.5  & 0.5  & 1.2 \\
 \hline
 {\bf H--2}  &$ -.04    $&
       208 &       200 &       208 &       209 &
      .574 &      .582 &      .563 &      .592 &
      24.8 &      23.3 &      24.4 &      25.9 \\
1.06      & $ -.07    $&
   6 &   5 &   5 &   6 &
.013 &.013 &.014 &.024 &
 1.2  & 0.8  & 0.5  & 1.8 \\
 \hline
 {\bf H--3}  &$ -.01    $&
       188 &       185 &       188 &       188 &
      .568 &      .573 &      .566 &      .561 &
      20.0 &      19.6 &      20.0 &      19.7 \\
0.99      & $ +.03    $&
  12 &   5 &  12 &  12 &
.020 &.012 &.033 &.028 &
 2.0  & 0.6  & 1.4  & 2.2 \\
 \hline
 {\bf H--4}  &$ +.02    $&
       197 &       199 &       198 &       196 &
      .789 &      .787 &      .790 &      .786 &
      30.7 &      31.3 &      30.9 &      30.2 \\
0.98      & $ +.02    $&
   4 &   4 &   3 &   5 &
.009 &.007 &.009 &.012 &
 1.2  & 1.1  & 0.7  & 1.7 \\
 \hline
 {\bf H--5}  &$ +.08^*  $&
       210 &       227 &       211 &       210 &
      .525 &      .526 &      .562 &      .525 &
      23.1 &      27.1 &      25.0 &      23.1 \\
0.92$^*$  & $ +.08^*  $&
   6 &   7 &   5 &   6 &
.015 &.016 &.014 &.031 &
 1.2  & 0.9  & 0.5  & 1.8 \\
 \hline
 {\bf H--6}  &$ +.00    $&
       193 &       190 &       192 &       190 &
      .440 &      .443 &      .440 &      .501 &
      16.3 &      16.1 &      16.1 &      18.1 \\
1.12      & $ -.22    $&
   6 &   6 &   7 &   6 &
.015 &.016 &.018 &.030 &
 0.9  & 0.5  & 0.5  & 1.3 \\
 \hline
 {\bf H--7}  &$ +.10    $&
       199 &       213 &       200 &       202 &
      .778 &      .762 &      .783 &      .786 &
      30.7 &      34.5 &      31.4 &      31.9 \\
1.02      & $ -.05    $&
   5 &   5 &   3 &   6 &
.009 &.008 &.009 &.013 &
 1.2  & 1.3  & 0.7  & 2.0 \\
 \hline
 {\bf Z--1}  &$ -.07    $&
       221 &       208 &       218 &       207 &
      .856 &      .865 &      .854 &      .836 &
      41.7 &      37.5 &      40.5 &      35.9 \\
0.89      & $ +.17    $&
  10 &   8 &  10 &  12 &
.011 &.008 &.018 &.019 &
 3.4  & 2.6  & 3.2  & 4.5 \\
 \hline
 {\bf Z--2}  &$ +.03    $&
       220 &       227 &       220 &       220 &
      .497 &      .490 &      .505 &      .507 &
      24.1 &      25.2 &      24.4 &      24.6 \\
1.00      & $ -.04    $&
  11 &   6 &  10 &  11 &
.019 &.010 &.025 &.034 &
 1.9  & 0.7  & 1.2  & 2.4 \\
 \hline
 {\bf Z--3}  &$ +.02    $&
       228 &       236 &       225 &       230 &
      .306 &      .299 &      .319 &      .299 &
      15.9 &      16.6 &      16.2 &      15.8 \\
0.97      & $ +.03    $&
  14 &  10 &  21 &  17 &
.023 &.017 &.040 &.042 &
 1.4  & 0.5  & 0.9  & 1.6 \\
 \hline
 {\bf Z--4}  &$ +.12^*  $&
       227 &       253 &       233 &       228 &
      .728 &      .721 &      .752 &      .731 &
      37.5 &      46.1 &      41.0 &      37.8 \\
0.92$^*$  & $ +.07^*  $&
   9 &   6 &  12 &  10 &
.014 &.008 &.021 &.021 &
 2.5  & 1.6  & 3.1  & 3.4 \\
 \hline
 {\bf Z--5}  &$ +.10    $&
       207 &       232 &       200 &       206 &
      .305 &      .285 &      .350 &      .310 &
      13.1 &      15.4 &      14.0 &      13.2 \\
0.94      & $ -.03    $&
  13 &  10 &  15 &  15 &
.024 &.017 &.033 &.044 &
 1.2  & 0.4  & 0.7  & 1.4 \\
 \hline
 {\bf Z--6}  &$ +.03    $&
       185 &       191 &       186 &       183 &
      .608 &      .592 &      .610 &      .591 &
      20.7 &      21.6 &      21.0 &      19.8 \\
0.95      & $ +.07    $&
  12 &  11 &  12 &  12 &
.023 &.028 &.054 &.054 &
 2.1  & 1.6  & 1.5  & 3.2 \\
 \hline

\end{tabular}
\caption  {\label{tab9}        
\sl Kinematic properties of the 13 events observed by the H1 and ZEUS
collaborations which have, at least in one method, 
$Q^2 > 15000$ GeV$^2$ and $M >$ 180 GeV. 
For event Z--6, only the DA values are
accurate, the values of the other methods being extrapolated.
The values of M, $y$ and $Q^2$ are identical to those of the original
papers ($e$, DA, $\SSS$  for H1~\cite{H1VHQ2,GREGQ2}, $e$, 
DA for ZEUS~\cite{ZEUSVHQ2}).
The new values of the table are for the $\SSS$ (ZEUS) and for the
$\omega$ (H1+ZEUS), and for the non available errors (see text for more 
details).
Also given are $\sigma_{he}$,\deel and \dssl, which allow to quantify
the size of the errors and to tag the presence of initial state radiation
and to correct for it (marked with a ``*'').
After ISR corrections, the ($\sigma_{he}$,\deel,\dssl) 
values are for event H--5: (1.00,.00,.00)
and for event Z--4: (1.00,+.04,--.01). 
The masses are given in GeV, the $Q^2$ in $10^3$ GeV$^2$.
}
\end{small}
\end{table}
\subsection{Mass Distribution}
In the following we will not consider any more the event Z--5, since it 
survives the $Q^2$ cut only  for the DA method.  It might be a radiative
event which could not be identified 
due to a large smearing in the hadronic energy.
In any case (radiative or not) it would not survive the $Q^2_{\omega}$ cut,
which is one of the 2 conditions the final sample must satisfy. Note 
``en passant'' that its mass is reduced from $M_{DA}$=232 GeV to 
$M_{\omega}$=207 GeV. The mass distributions of the remaining 12 events,
 obtained with the
$e,$ DA, $\SSS$ and $\omega$, are displayed in fig.5a,b,d,e.
\begin{figure}[htb]                                                           
\begin{center}                                                                 
\epsfig{file=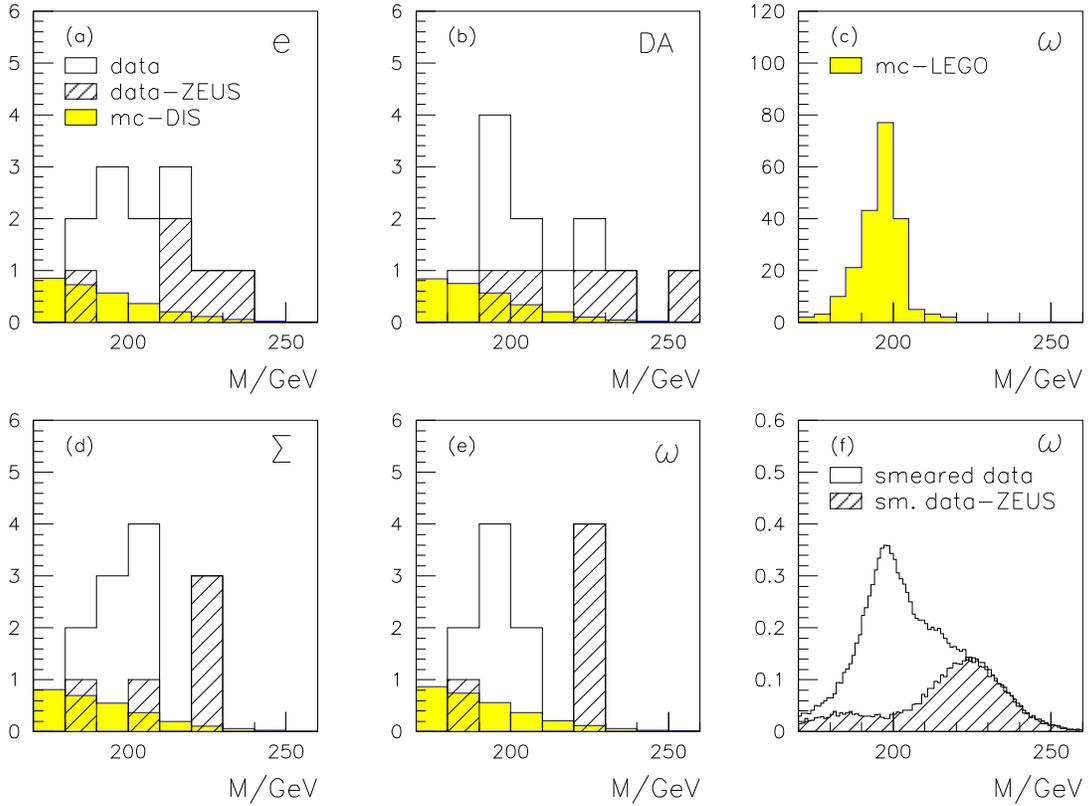,width=15cm,
  bbllx=50pt,bblly=230pt,bburx=530pt,bbury=585pt} 
\end{center}                                                                   
\caption[]{\label{fig.8}
\sl {
Mass distributions obtained with 
 the $e$ (5a), DA (5b),  $\SSS$ (5d) and  $\omega$ (5e) reconstruction methods.
The main histogram represents the mass distribution of the H1 and ZEUS events
at $Q^2 > $ 15000 GeV$^2$ and $M > $ 180 GeV$^2$.
The dashed histogram represents
the contribution of the ZEUS events only.
The grey histogram represents the prediction from Standard Model DIS, obtained
from a simulation normalized to the summed luminosity of H1 and ZEUS.  
Fig.5c displays the $\omega$ mass reconstruction of a 200 GeV narrow resonance
generated with the LEGO program~\cite{lego} and 
fully simulated in the H1 detector. Fig.5f is obtained from
fig.5e by filling one  gaussian distribution per event, 
with  mean and width given by
its $\omega$-mass and its error (cf table 1). }
}
\end{figure}        

\newpage

\vspace*{0.5cm}

The event Z--4  has the highest mass at $253\pm6$ GeV. It is identified as
a radiative event and its mass is reduced  by
the $\omega$ determination to $226\pm 5$ GeV, thereby
reducing the scattering appearance of the ZEUS events.
Note that the error quoted on the DA mass is much smaller than the effect
due to radiation.

For the 5  ZEUS events 
the weighted average mass is decreased from M$^{avg}_{DA}$=226$\pm 9$ GeV 
 to  M$^{avg}_{\omega}$=216$\pm 7$  GeV.
Actually with the $\omega$ determination 
 4 out of the 5 ZEUS events  lie between 220 and 228 GeV, the 5$^{th}$ one
(Z--6) being at 185 GeV, i.e. at a lower mass than any of the 7  H1 events
(see fig.5b).

The average mass\footnote{
The fact that the average masses are slightly different from the original
publications is due to the difference in the errors which weights
the events in a different way. This difference is irrelevant in the
current discussion.} 
of the 7 H1 events is essentially not dependent on the method
used:  M$^{avg}_{\omega}$=199$\pm 2.5$ GeV, 
M$^{avg}_{e}$=200$\pm 2.6$ GeV,  M$^{avg}_{DA}$=201$\pm 5$ GeV.
The 7 H1 events remain clustered   between 188 and 211 GeV. Thus the H1 and
ZEUS  
samples are concentrated  at significantly different mass values and 
this splitting cannot
be accounted for either by ISR or by detector effects. 

The small number of events involved 
prevents a definite interpretation of this effect. However the fact that 
no event among  the 5 of ZEUS is found in the bin where the 7 H1 events
are measured  suggests that this specific accumulation is
a statistical fluctuation. In fig.5c is shown the reconstruction of a 
narrow resonance generated at M=200 GeV with the LEGO~\cite{lego}
program in the H1 detector using the $\omega$ 
method\footnote{All the methods give similar distributions on these events,
except for the DA which has a larger r.m.s. due to radiative tails.
In  units of GeV the (mean; r.m.s.) are for 
the $\omega$, $e$, DA and $\SSS$ respectively:
(195.5; 7.5), (195.3; 7.4),  (195.2; 9.6), (195.3; 7.8). The bias of 
about 5 GeV can be removed by taking into account the mass of the jet,
but is present when using inclusive methods.}.
The width
of the distribution which include experimental and QED/QCD radiation
effects,
 cannot accomodate the tails of the measured experimental 
distributions (fig.5a,b,d,e). 

If we make an ideogram from the histogram 5e
i.e. if  we apply gaussian smearing to each of the 12 events
using their $\omega$ mass as a mean, and their error as r.m.s., we obtain
fig.5f, which also shows the incompatibility with fig.5c.
To reconcile the 2 ``peaks'' visible in fig.5f would
require to miscalibrate uniformly the electron energy of
the H1 events at least by +6\% and the ZEUS events by $-6$\%, values completely
incompatible with the absolute scale uncertainties
  found in the previous section. Note however, that the H1 events alone
support the narrow resonance hypothesis.

Since after the $\omega$ kinematic treatment 
none of the 12 events migrates outside  the
very high $Q^2$ and mass region, we   confirm  that
the visible excess
published by the collaborations at very high $Q^2$ is not due
to detector (calibration) or radiation (ISR) effects.
        
\vspace*{.5cm}

\section{Conclusion}

We have introduced a new reconstruction  method  (``$\omega$'') which
allows  the kinematic variables of the high $Q^2$
events to be determined in a more precise way. It uses 
the kinematic constraints on an event by event basis to calibrate
the hadronic energy
and to identify  and correct for the presence of QED initial state radiation.

This method has been applied to the 12 HERA events observed by the H1 and
ZEUS collaborations 
(for an expectation of about 5) at $Q^2>15000$ GeV$^2$ and M$>180$ GeV.
The accumulation of the 7 H1 events around 200$\pm12.5$ GeV is confirmed.
The average mass of the 5 ZEUS events is decreased from the double-angle
value of 226$\pm 9$ GeV to 
216$\pm 7$ GeV.
 However none of the ZEUS events enter the H1 accumulation region.
Taking the estimation of the systematic errors at face value, 
this suggests that, if the observed excess at very high $Q^2$
is due to physics beyond the
Standard Model,
it is unlikely to be explained by the decay of a single narrow  resonance
such as a leptoquark.
The increase of the integrated 
luminosity in the current and in the future years at
HERA will clarify the origin of this effect.

\vspace*{0.5cm}

\begin{Large}      
\noindent
{\bf Acknowledgments}
\end{Large}
\vspace*{0.4cm}

We would like to thank the two collaborations for the data we have studied
in this paper. In particular this work has taken place within the
H1 collaboration, and some of the results obtained were the outcome of
the efforts of many people of the ``Beyond the Standard Model''  group 
to get the analysis of the very High $Q^2$ events completed.
We also would like to thank J. Dainton, R. Eichler, J. Gayler, D. Haidt,
B. Straub and G. Wolf for a 
careful reading of the manuscript and for their useful remarks.

\vspace*{0.7cm}
\begin{Large}      
\noindent
{\bf Appendix:} \hspace*{0.7cm}    {\large $y$ and $Q^2$  formulae
 ($M\equiv \sqrt{Q^2/y}$)}   
\end{Large}

\vspace*{0.5cm}   
\begin{center}
\begin{tabular}{|c|c|c|}
\hline  
method &  $y$ & $Q^2$ \htab\\
\hline  
 $ e $ &  $1-${\large$\frac{E}{E_{\circ}}$}$\sin^{2}$\thee &
 {\large $\frac{p_{T,e}^2}{1-y_{e}}$ }     \hhtab \\

\hline      
 $ h $    &  {\large $\frac{\Sigma}{2 E_{\circ}} $}
          &  {\large $\frac{p_{T,h}^2     }{1-y_h }$}   
          \hhtab\\ 
\hline  
  DA    &  {\large $\frac{\tan{\frac{\gamma}{2}}}
{\tan{\frac{\gamma}{2}}+\tan{\frac{\theta}{2}}} $}        
          &  $4E^{2}_{\circ} ${\large$\frac{\cot{\frac{\theta}{2}}}
{\tan{\frac{\gamma}{2}}+\tan{\frac{\theta}{2}}}$}
          \hhtab \\     
\hline
 $\Sigma$ &  {\large $\frac{\Sigma}{\Sigma+E(1-\cos\theta)}$}     
          &  {\large $\frac{p_{T,e}^2}{1-y_{\Sigma}}$ }     
          \hhtab\\
\hline
\end{tabular}
\end{center}

\end{document}